\documentclass[12pt,a4paper]{article}
\usepackage{a4,epsf,here}
\def\trM{{\rm tr}\,M_{\rm red}}
\def\be{\begin{equation}}
\def\ee{\end{equation}}
\def\bea{\begin{eqnarray}}
\def\eea{\end{eqnarray}}
\def\eq#1{(\ref{#1})}
\def\bs{\bigskip}
\def\ms{\medskip}
\def\hom{\hbar\omega}
\def\fig#1{figure \ref{#1}}
\def\tab#1{table \ref{#1}}

\def\siml{\,\hbox{\kern.1em \lower.6ex \hbox{$\sim$} \kern-1.12em
          \raise.6ex \hbox{$<$} }}
\def\simg{\,\hbox{\kern.1em \lower.6ex \hbox{$\sim$} \kern-1.12em
          \raise.6ex \hbox{$>$} }}
\def\d{{\rm d}}

\newcommand{\Figurebb}[9]{
\begin{figure}[H]
\leavevmode
\epsfysize=#7cm
\epsfbox[#2 #3 #4 #5]{#6}
\par
\parbox{#8cm}{
\caption[figure]{\renewcommand{\baselinestretch}{0.8} \small
                                           \hspace{-0.3truecm}#9}
\label{#1}}
\end{figure}
}

\newcommand{\Table}[4]{
\begin{table}[H]\begin{center}{#3}
\parbox{#2cm}{
\vspace{0.5cm}
\caption[table]{\renewcommand{\baselinestretch}{0.8} \small
                                           \hspace{-0.3truecm}#4}
\label{#1}}
\end{center}
\end{table}
}

\def\H1{\widehat{H}_1}

\renewcommand{\d}{\ensuremath{\mathrm{d}}}

\def\alstab{\chi}

\begin{document}

\baselineskip 14pt

\centerline{\bf \Large On the canonically invariant calculation 
                of Maslov indices}

\bs
\bs

\centerline{\bf M Pletyukhov$^1$ and M Brack}

\bs

{\small
\centerline{Institute for Theoretical Physics, University of
Regensburg, D-93040 Regensburg, Germany}
\centerline{$^1$e-mail: mikhail.pletyukhov@physik.uni-regensburg.de}
}
 
\bs
\ms

\noindent 
{\bf \large Abstract} 

\ms
\noindent
{\small
After a short review of various ways to calculate the Maslov index 
appearing in semiclassical Gutzwiller type trace formulae, we discuss a 
coordinate-independent and canonically invariant formulation recently 
proposed by A Sugita (2000, 2001). We give explicit formulae 
for its ingredients and test them numerically for periodic orbits in 
several Hamiltonian systems with mixed dynamics. We demonstrate how
the Maslov indices and their ingredients can be useful in the
classification of periodic orbits in complicated bifurcation scenarios,
for instance in a novel sequence of seven orbits born out of a tangent 
bifurcation in the H\'enon-Heiles system.
}

\section{Introduction}

The idea of Gutzwiller \cite{gutz} to express the quantum-mechanical
density of states of a Hamiltonian system in terms of its classical
periodic orbits and their properties has brought an enormous progress 
to the field of ``quantum chaos'' \cite{chaos}. The trace formula
which he derived in 1971 \cite{gutz} is valid for systems in which all 
periodic orbits are isolated in phase space. Later versions of trace 
formulae were formulated for integrable systems \cite{bablo,beta} and 
mixed systems with continuous symmetries \cite{bablo,stma,crli}, and 
very recently also for systems with explicit spin degrees of freedom 
\cite{boke,plet}.

An important ingredient in all these trace formulae is, besides the
action and stability of a periodic orbit, the so-called Maslov index. 
It is an invariant property of a periodic orbit which can change only 
when the orbit undergoes a bifurcation or when a continuous symmetry 
is broken or restored under the variation of a system parameter (eg 
energy, deformation or an external field). The calculation of the
Maslov index is not always straightforward, in particular for systems
with many degrees of freedom or systems which are not of the ``kinetic
plus potential energy'' type. In the standard methods used in the
literature \cite{liro,wint,crliro}, the determination of the Maslov 
index of a stable orbit necessitates the explicit use of an ``intrinsic'' 
coordinate system  that follows the orbit (as introduced by Gutzwiller 
\cite{gutz}), which can be numerically quite cumbersome. Easy-to-use 
calculational recipes using the method of \cite{crliro} have been given
in the appendix D of \cite{book}.

Recently, Sugita \cite{sugita} has given a formula for the Maslov index 
which only contains canonically invariant ingredients. However, no 
practical recipes were given for the explicit calculation of the winding
number $m$ [see equation \eq{main} in section \ref{secmain}]. Inspired by 
another recent work on semiclassical trace formulae and Maslov indices 
\cite{murat}, we have developed simple calculational recipes for the 
winding number and the stability angle used in Sugita's formula \cite{sugita}. 
The purpose of the present paper is to implement these recipes 
for various Hamiltonian systems with mixed dynamics and to test their 
results towards those obtained with the standard methods 
\cite{gutz,wint,crliro}. As an outcome, we shall give some empirical 
rules for the behaviour of some of the ingredients of the Maslov indices
at bifurcations and demonstrate that they can be useful for the 
classification of periodic orbits in connection with complicated 
bifurcation scenarios.

\section{General formalism}
\label{secgen}

\subsection{Trace formulae and role of Maslov indices}

For isolated orbits, Gutzwiller's trace formula \cite{gutz} for the 
oscillating part of the density of states $g(E)$ reads
\be
\delta g (E) = \frac{1}{\pi \hbar}\,\sum_{\mathrm{ppo}}\,T_{\mathrm{ppo}}(E)\,
               \sum_{r=1}^{\infty} 
               \frac{1}{\sqrt{|\det(M^r_{\mathrm{red}}-I)|}}\,
               \cos\left[\frac{r}{\hbar}\,S_{\mathrm{ppo}}(E)
                         -\frac{\pi}{2}\,\sigma_r \right],
\label{tfiso}
\ee
where $T_{\mathrm{ppo}}$ and $S_{\mathrm{ppo}}$ are the periods and actions 
of the primitive periodic orbits (ppo) and $r$ is an index which counts the
repetitions of the primitive orbits. $M_{\mathrm{red}}$ is their reduced 
monodromy matrix -- often called stability matrix -- and $\sigma_r$ 
is the Maslov index of the repeated orbit (which for stable orbits is {\it not}
a multiple of that of the primitive orbit). For isolated orbits the Maslov
index is always an integer, irrespectively of the dimensionality of the 
system.

For systems with continuous symmetries, periodic orbits are no longer
isolated in phase space but come in degenerate families. The semiclassical
trace formulae can still be written in the general form
\be
\delta g (E) = \sum_{\mathrm{po}}\, {\cal A}_{\mathrm{po}}(E)\,
               \cos\left[\frac{1}{\hbar}\,S_{\mathrm{po}}(E)
                    -\frac{\pi}{2}\,\sigma_{\mathrm{po}}\right],
\label{tfgen}
\ee
where ``po'' now refers to all repeated families {\it and} isolated orbits. 
The amplitudes ${\cal A}_{\mathrm{po}}(E)$ depend on the degeneracies of 
the orbit families, besides their stabilities, and have been given in the 
literature \cite{bablo,stma,crli}. The Maslov index $\sigma_{\mathrm{po}}$ 
of an orbit family can be a half integer. Its determination, especially 
for families with high degeneracies such as they occur in isotropic 
harmonic oscillators with SU(N) symmetry, is by no means trivial. Although 
some hints may be found in \cite{crli,sugita,murat,long} there exists, to our 
knowledge, no simple recipe for their calculation. We shall in the 
remainder of the present paper restrict ourselves to systems with isolated 
orbits.

It has been shown \cite{crliro,wnrobb} that the Maslov index of an isolated
orbit is a canonical invariant, independent of the orbit's intrinsic 
coordinate system and of the starting point for the calculation of the 
monodromy matrix. However, the published practical ways for its calculation 
\cite{gutz,wint,crliro} do make explicit use of the orbit's intrinsic 
coordinate system. Furthermore, using the form
\be
\sigma_{\rm po}=\mu_{\rm po}+\nu_{\rm po}                                     
\label{munu}
\ee
given in \cite{crliro} -- whereby the part $\mu_{\rm po}$ is the Morse index
appearing in the semiclassical Green's function and the part $\nu_{\rm po}$ 
comes from the stationary-phase integrations transverse to the orbit -- it
has been explicitly demonstrated for an unstable orbit that $\mu$ and $\nu$ 
alone need not be invariant and may, in fact, depend explicitly on its 
starting point.

A remark on the name ``Maslov index'' might be in place here. Originally, 
the Maslov index was introduced in the framework of the WKB and later the 
EBK quantization of integrable systems \cite{kelrub,maslov,arnold}. The 
quantity $\sigma_{\rm po}$ appearing in the above trace formulae is a 
different index, although for integrable systems it can be easily related 
\cite{beta,crli} to the Maslov indices of the EBK quantization. In the 
standard literature on the periodic orbit theory, the name of  
$\sigma_{\rm po}$ has, however, established itself as ``Maslov index'' 
and we want here to adhere to this convention. 

The trace formulae \eq{tfiso} and \eq{tfgen} provide an approximative
description of the quantum-mechanical density of states in terms of 
classical periodic orbits and their properties. It is obvious that the
Maslov index $\sigma_{\rm po}$ plays a vital role in establishing the
correct quantum phase interferences and therefore must be determined
correctly. 

\subsection{Invariant calculation of the Maslov index}
\label{secmain}

All information about the Maslov index of an isolated orbit is contained
in its matrizant $M(t)$ describing the time propagation of a small
perturbation $\delta q,\delta p$ around the orbit in phase space:
\be
\left( \begin{array}{c} \delta q(t)\\ \delta p(t) \end{array} \right)
      \; = \; M(t)
\left( \begin{array}{c} \delta q(0)\\ \delta p(0) \end{array} \right),
\label{pert}
\ee
where $q(t)$ and $p(t)$ are $N$-dimensional coordinate and momentum
variables. $M(t)$ is obtained by solving the linearized equations of 
motion of a classical system characterized by its Hamiltonian $H(q,p)$, 
which leads to the differential equations
\be
\frac{\d}{\d t} M(t) = J H''(t) M(t) , \qquad M(0)=I_{2N}\,,
\label{matriz}
\ee
where
\be
J = \left( \begin{array}{cc}
0 & I_N \\
\!-I_N & 0 \end{array} \right)  , \qquad
H''(t) = \left( \begin{array}{c@{\hspace{.1cm}}c} 
  \frac{\partial^2 H}{\partial q\partial q} 
& \frac{\partial^2 H}{\partial q\partial p} \vspace*{0.15cm} \\
  \frac{\partial^2 H}{\partial p\partial q} 
& \frac{\partial^2 H}{\partial p\partial p}
\end{array} \right).
\ee
$J$ is the unit symplectic matrix in the $2N$-dimensional phase space, 
and $I_{2N}$, $I_N$ are the $2N$- and $N$-dimensional unit matrices,
respectively. At the time of the orbit's period $T$, the matrix $M(T)$ is 
called the monodromy matrix. One can always transform $M(T)$ into block 
form. One parabolic (2$\times$2) block contains the  trivial unit 
eigenvalues related to energy conservation; the remaining 
$(2N-2)$-dimensional part of $M(T)$ is called the reduced monodromy matrix 
$M_{\rm red}$ or stability matrix. $M_{\rm red}$ falls into (inverse) 
parabolic, elliptic, or (inverse) hyperbolic (2$\times$2) blocks or, for 
$N>2$ dimensions, loxodromic  (4$\times$4) blocks, depending on the 
stability of the respective orbits.

The amplitude in the Gutzwiller trace formula \eq{tfiso} diverges when
$\det(M^r_{\mathrm{red}}-I)$ becomes zero. For isolated orbits this 
happens at bifurcations, where an orbit changes from stable to 
hyperbolically unstable or vice versa, or when a continuous symmetry is 
restored under the variation of a system parameter such as energy, 
deformation or some external field. At such singular points 
$M^r_{\mathrm{red}}$ contains an extra parabolic block. The trace formula 
\eq{tfiso} then cannot be used and one must resort to uniform approximations 
\cite{ozohan,ssun,toms,hhun} which we, however, shall not be concerned 
with in the present paper. We just mention here that bifurcation and 
symmetry-restoring points are the only places where a Maslov index is
allowed to change. The corresponding rules for bifurcations can be found 
in \cite{ssun}; an example for the changes of Maslov indices under symmetry 
breaking (or restoring) will be mentioned in section \ref{secho}.

Referring to the block form of the stability matrix, Sugita \cite{sugita} 
has recently proposed the following general formula for the Maslov index 
$\sigma_r$ in \eq{tfiso} of an isolated periodic orbit: 
\be
\sigma_r = \sum_{i=1}^{n_{{\rm ell}}} 
           \left(1 + 2 \left[ \frac{r \alstab_i}{2 \pi} \right] \right)
           + r n_{{\rm ih}} + 2 mr\,.                        \label{main} 
\ee
Here $[x]$ denotes the integer part of $x$ (ie the largest integer number 
$\leq x$), $n_{{\rm ell}}$ is the number of elliptic blocks and $\chi_i$ the 
stability angle of the $i$th elliptic block, $n_{{\rm ih}}$ is the number 
of inverse-hyperbolic blocks, and $m$ is a winding number. Sugita has also 
discussed in \cite{sugita} the relation of the winding number $m$ to the 
homotopy theory. However, no explicit prescription for the computation of
$m$ has been given.

This question has been partially answered in a recent review article on 
periodic orbit theory \cite{murat}. The winding number $m$ has been identified 
as the Gel'fand-Lidski winding number \cite{gellid}, and it has been explained 
how to calculate it in principle. Muratore-Ginanneschi \cite{murat} has also 
discussed a relation of $\sigma_r$ given by (\ref{main}) to an index 
which is known in the mathematical literature after the names of Conley and 
Zehnder \cite{conl}, and has provided extensive references on the latter 
subject. However, in \cite{murat} the way of extracting a unique value of the 
stability angle $\chi_i$ from the the eigenvalues $e^{\pm i \chi_i}$ of the 
stability matrix has only been hinted at, and a practical algorithm still 
remained to be specified.
We also want to remark that a representation similar to (\ref{main}) 
appears in a mathematical paper \cite{long} where a classification 
of the admissible normal forms of the elements of Sp(2N) is given.

The goal of the present section is to specify the definitions of $\chi_i$ 
and $m$, in order to make the definition of $\sigma_r$ in (\ref{main}) unique 
and useful for practical applications. In particular, we are going to 
establish how the choice of $\chi_i$ made in \cite{sugita} corresponds to 
the prescription for the calculation of $m$ given in \cite{murat}. 

Following \cite{murat}, we split $M(t)$ into a product of a periodic and 
an average part (also called the Floquet decomposition):
\be
M(t) = M_{{\rm per}}(t) M_{{\rm av}} (t)                   \label{floq}
\ee
with
\be
M_{{\rm av}} (t) = \exp (t K)\,,
\ee
where $K$ is a constant matrix. By definition, the periodic part of the 
matrizant in (\ref{floq}) satisfies the condition $M_{{\rm per}}(t)=
M_{{\rm per}}(t+T)$. In particular, $M_{{\rm per}}(0)=M_{{\rm per}}(T)=I_{2N}$.
 We therefore can specify the constant matrix $K$ by equating
\be
M_{{\rm av}} (T) \equiv \exp (T K) = M(T)\,.
\label{jhav}
\ee
Then, we get
\be
K = \frac{1}{T} \ln [M(T)]\,.
\label{logar}
\ee
To take the logarithm on the rhs of (\ref{logar}), we diagonalize $M(T)$, 
calculate the logarithms of the eigenvalues of $M(T)$, and then return to 
the initial basis. However, the relation (\ref{logar}) remains symbolic 
until we adopt a certain phase convention for determining the eigenvalues 
of $K$.  

In the standard definition of the function $\ln(z)$ with $z=|z|e^{i\phi}$, 
the phase range $\phi\in(-\pi,\pi]$ is chosen, corresponding to the branch 
cut line being taken along the negative real axis. Let us consider the 
eigenvalue problem
\be
M(T)\, \xi_i^{\pm} = e^{\pm i \widetilde{\chi}_i}\, \xi_i^{\pm}\,, 
\ee
with $\xi_i^{-} = [\xi_i^{+}]^*$ and
\be
\widetilde{\chi}_i = - i \ln [e^{+ i \widetilde{\chi}_i }] \in (0, \pi)\,.
\ee
The case $\widetilde{\chi}_i = \pi$ will be discussed separately below.

Let us now introduce the symplectic product 
\be
s_i=  + [ {\rm Re} (\xi_i^{+})]^{{\rm T}} J \, {\rm Im} (\xi_i^{+}) \equiv 
-  [ {\rm Re} (\xi_i^{-})]^{{\rm T}} J \, {\rm Im} (\xi_i^{-})  \, ,
\label{sign}
\ee
known as the Krein invariant \cite{arnold,krein}. With this, 
we can adopt the following convention for the eigenvalues $\pm i 
\frac{\chi_i}{T}$ of $K$:
\be
K \xi_i^{\pm} = \pm i \frac{\chi_i}{T}\, \xi_i^{\pm}\,,
\label{defk}
\ee
so that 
\bea
\chi_i &=& \widetilde{\chi}_i  \qquad \quad\,\, {\rm for} \quad s_i > 0\,, 
           \nonumber\\
\chi_i &=& 2 \pi -\widetilde{\chi}_i  \quad {\rm for} \quad s_i < 0\,. 
\eea
These relations fully determine the constant matrix $K$ and specify uniquely
the stability angle to be used in the formula (\ref{main}). In this phase
convention, $\chi_i$ takes values in the range $(0, 2 \pi)$. 

The case of an inverse parabolic block with $\,e^{\pm i \widetilde{\chi}_i }
=-1$ is degenerate and requires special consideration. It occurs when 
$\mathrm{tr} M_{\mathrm{red}} = -2$ and corresponds to the stability 
changing between elliptic and inverse hyperbolic. In this case we choose 
the value $\chi_i = \pi$ by continuity reasons. The inverse parabolic 
block should be taken into accout in the formula (\ref{main}) as a 
special case of either an inverse hyperbolic or an elliptic block, but not 
twice -- in order to avoid double counting.

The winding number $m$ is an invariant characteristic of $M_{{\rm per}}(t) 
=M(t)M^{-1}_{{\rm av}}(t)$. To determine it, it is convenient to employ the 
so-called polar decomposition of the symplectic matrix $M_{\mathrm{per}}$ 
into a product of an orthogonal matrix $R_{\mathrm{per}}$ and a 
positive-definite symmetric matrix $W_{\mathrm{per}}$: 
\be
M_{\mathrm{per}} = R_{\mathrm{per}} \, W_{\mathrm{per}}\,.    
\label{poldec}
\ee
In turn, the orthogonal matrix $R_{\mathrm{per}}$ admits the representation
\be
R_{\mathrm{per}} = \left( \begin{array}{cc} X_{\mathrm{per}} 
                 & Y_{\mathrm{per}} \\ -Y_{\mathrm{per}} 
                 & X_{\mathrm{per}} \end{array} \right).   \label{orthog}
\ee
Therefore, the winding number $m$ can be defined as 
\be
m= \varphi(T) - \varphi(0)\,,                             \label{windnum}
\ee
where
\be
\varphi (t) = \frac{1}{2 \pi} \mathrm{Arg} \det \left[ X_{{\rm per}} (t) 
              + i Y_{{\rm per}} (t)\right].
\label{phit}
\ee
Since $X_{\mathrm{per}} (t)$ and $Y_{\mathrm{per}} (t)$ are periodic, 
$m$ is a (positive or negative) integer number. 

The winding number (\ref{windnum}) has been vastly discussed in the 
literature, both mathematical and physical (see \cite{murat} for 
extensive references). In particular, we would like to quote here 
that it has been introduced in \cite{gellid} for a topological 
characterization of the structural stability of linear Hamiltonian flow. 

The extraction of $R_{{\rm per}}(t)$ from $M_{{\rm per}}(t)$ provides a 
nice representation of the evolution of $\varphi(t)$, because $\det 
[X_{{\rm per}}(t)+iY_{{\rm per}}(t)]$ runs around the unit circle. However, 
the polar decomposition (\ref{poldec}) is not essential for the calculation 
of the winding number $m$, even though the latter is encoded in 
$R_{\mathrm{per}}(t)$. The same result as in (\ref{phit}) can be also 
obtained from $\psi(T)-\psi(0)$, where
\be
\psi (t) = \frac{1}{2 \pi} \mathrm{Arg} \det \left[ A_{{\rm per}} (t) 
                      + i B_{{\rm per}} (t)\right],
\ee 
and the matrices $A_{{\rm per}}(t)$ and $B_{{\rm per}}(t)$ are the blocks of
\be
M_{{\rm per}} (t) = \left( \begin{array}{cc} A_{{\rm per}} (t) & B_{{\rm per}} 
                    (t) \\ C_{{\rm per}} (t) & D_{{\rm per}} (t) \end{array} 
                    \right).
\ee
For a proof and further discussion of this point, see the Appendix A of 
\cite{littrep}.

Let us now consider the following canonical transformation  
\be
M (t)= S(t) M_{{\rm av}} (t) S(0)^{-1}\,.
\ee  
with $S(t) \equiv M_{{\rm per}} (t)$. Since $S(0)= I_{2N}$, this expression is 
equivalent to (\ref{floq}). The  relation between the Maslov indices $\sigma$ 
of $M(t)$ and $\sigma_{{\rm av}}$ of $M_{{\rm av}}(t)$ for $r=1$ is given by 
\cite{sugita}
\be
\sigma = \sigma_{{\rm av}} + 2m\,.                          \label{sugtheo}
\ee
But the winding number in $\sigma_{{\rm av}}$ equals zero, since 
$M_{{\rm av}}(t)$ belongs to the same homotopy class as the identity 
matrix, ie it can be continuously shrunk to the latter. Therefore, 
in order to determine $\sigma_{{\rm av}}$ we just need to find the 
number of elliptic and inverse hyperbolic blocks, which can be read off 
the block form of $M_{{\rm av}}(T)=M(T)$. We remark that neither 
$\sigma_{{\rm av}}$ nor $m$ depend on the choice of the starting point 
on the periodic orbit \cite{sugita}, as it must be for canonically
invariant quantities. 

We note in passing that the Maslov index $\sigma_r$ can be identified 
with the winding number obtained from a polar decomposition of the whole 
matrizant $M(t)$ \cite{wnrobb}. However, that approach also requires some 
further specifications for stable orbits.

This completes our specification of Sugita's approach. Before discussing
another choice of the phase convention which allows us to make contact with 
the earlier approach of \cite{crliro}, we will illustrate our method with an 
analytical example.

\subsection{Analytical example: irrational harmonic oscillators}
\label{secho}

We consider here a simple integrable system with isolated orbits, for 
which all the above quantitites can be evaluated analytically. This
is the two-dimensional anisotropic harmonic oscillator
\be
H = \frac12\, (p_x^2 + p_y^2) 
  + \frac12 \,(\omega_x^2 x^2 + \omega_y^2 y^2 )
  = \frac{\omega_x}{2}\, (P_x^2 + Q_x^2) 
  + \frac{\omega_y}{2}\, (P_y^2 + Q_y^2)\,,
\ee
where $Q_x=x\sqrt{\omega_x}$, $Q_y= y\sqrt{\omega_y}$ and 
$P_x=p_x/\!\sqrt{\omega_x}$, $P_y=p_y/\!\sqrt{\omega_y}$. We assume that 
the frequencies $\omega_x$ and $\omega_y$ are incommensurate, so that the 
only periodic orbits are librations along the $x$ and $y$ axes; they 
are isolated and stable. For the orbit along the $x$ axis, the period 
is $T_x = 2 \pi / \omega_x$, and the monodromy matrix and its periodic 
and average parts are, respectively, given by (cf \cite{book,brjain})
\bea
M_x(t) &=& \left( \begin{array}{cccc}
\cos (\omega_x t) & 0 & \sin (\omega_x t) & 0 \\
0 & \cos (\omega_y t) & 0 & \sin (\omega_y t) \\
-\sin (\omega_x t) & 0 & \cos (\omega_x t) & 0 \\
0 & -\sin (\omega_y t) & 0 & \cos (\omega_y t) 
\end{array}\right) , \\
M_{x,{\rm per}} (t) &=& \left( \begin{array}{cccc}
\cos (\omega_x t) & 0 & \sin (\omega_x t) & 0 \\
0 & \cos (\Delta\omega_y \, t) & 0 & \sin (\Delta \, \omega_y t) \\
-\sin (\omega_x t) & 0 & \cos (\omega_x t) & 0 \\
0 & -\sin (\Delta\omega_y \, t) & 0 & \cos (\Delta \, \omega_y t) 
\end{array}\right) , \\
M_{x,{\rm av}} (t) &=& \left( \begin{array}{cccc}
1 & 0 & 0 & 0 \\
0 & \cos (\bar{\omega}_y t) & 0 & \sin (\bar{\omega}_y t) \\
0 & 0 & 1 & 0 \\
0 & -\sin (\bar{\omega}_y t) & 0 & \cos (\bar{\omega}_y t) 
\end{array}\right) ,
\eea
where 
\be
\bar{\omega}_y = \omega_y - \omega_x \left[
\frac{\omega_y}{\omega_x} \right] , \quad
\Delta \omega_y = \omega_y 
-\bar{\omega}_y = \omega_x \left[
\frac{\omega_y}{\omega_x} \right] .
\ee
In order to calculate the winding number $m_x$, we consider
\be
\mathrm{Arg} \det \left[X_{x,{\rm per}} (t) + i Y_{x,{\rm per}} (t)\right] 
= (\omega_x + \Delta \omega_y)\,t \,.
\ee
Then we obtain easily
\be
m_x = \frac{\omega_x + \Delta \omega_y}{\omega_x} = 1 + \left[
\frac{\omega_y}{\omega_x} \right].
\label{howind}
\ee
Next we calculate $\chi_x/2\pi$:
\be
\frac{\chi_x}{2 \pi} = \frac{\bar{\omega}_y}{\omega_x} =
\frac{\omega_y}{\omega_x} - \left[ \frac{\omega_y}{\omega_x} \right].
\label{hochi}
\ee
Finally, we obtain the Maslov index for the $r$th repetition to be
\be
\sigma_{x,r} = 1 + 2 \left[ \frac{r \chi_x}{2 \pi} \right] + 2rm_x 
= 1 + 2 \left[ r \frac{\omega_y}{\omega_x} - r \left[
\frac{\omega_y}{\omega_x} \right] \right] + 2r \left( 1 + \left[
\frac{\omega_y}{\omega_x} \right]\right) 
= 1 + 2r + 2 \left[ r \frac{\omega_y}{\omega_x} \right].
\label{anisharm}
\ee
This result agrees with that obtained in \cite{brjain} using the
method of \cite{crliro}.

A note about the isotropic harmonic oscillator with $\omega_x=
\omega_y=\omega$ may be of some interest here. The periodic orbits in this 
system, due to its SU(2) symmetry, are not isolated but form families of 
two-fold degenerate orbits. (The same is true for arbitrary rational axis 
ratios $\omega_x:\omega_y=n:p$ with integer $n,p$ for which the orbits are 
Lissajous figures.) As mentioned in the introduction, we lack a general
prescription for the calculation of the Maslov index of these families.
However, the semiclassical trace formulae of isotropic harmonic 
oscillators can be obtained by other means and are found to be
quantum-mechanically exact \cite{brjain}. In two dimensions, the trace
formula reads 
\be
g(E) = \frac{E}{(\hom)^2}\left\{ 1 + 2\sum_{r=1}^\infty
       \cos\left(r\frac{2\pi E}{\hom}\right)\right\},         
\label{ghoiso}
\ee
which supports a Maslov index $\sigma_r= 0$ (mod 4). Indeed, with our
above results we find $m=2$ from \eq{howind} and $\chi=0$ from \eq{hochi}, 
leading to $\sigma_r^{(0)}=4r$ which is equivalent to 0 (mod 4). The 
reason for our identifying this Maslov index for the isotropic harmonic 
oscillator here is that in the context of perturbation theory, trace 
formulae for slightly perturbed harmonic oscillators have been developed 
\cite{crpe,pert} in which the Maslov indices of the perturbed isolated 
orbits are obtained analytically, once the value $\sigma_r^{(0)}$ for the 
unperturbed families is known. For instance, in the H\'enon-Heiles system 
discussed later in section \ref{sechh}, the unperturbed families break up 
into three isolated orbits A, B and C as soon as the nonlinearity is turned 
on [$\varepsilon>0$ in \eq{hamhh}]. The changes in their Maslov indices with 
respect to $\sigma_r^{(0)}$ were found analytically \cite{hhun} to be 
$\Delta\sigma_A=+1$, $\Delta\sigma_B=0$ and $\Delta\sigma_C=-1$. Indeed, 
the numerical methods for the isolated orbits yield $\sigma_A=5$, 
$\sigma_B=4$ and $\sigma_C=3$, both using the method of \cite{crliro} 
(see \cite{hhpo}) and with our present method (see \tab{hhorb} below). 

A straightforward generalization for the $N$-dimensional harmonic
oscillator with irrational frequency ratios $\omega_i/\omega_j$ 
($i,j=1,2,\dots, N$) yields the Maslov index for the orbit along the 
$j$ axis
\be
\sigma^{(N)}_{j,r} = (N-1) + 2 r + 2 \sum_{{i=1}\atop{i\neq j}}^N 
                 \left[ r \frac{\omega_i}{\omega_j} \right] .
\ee

\subsection{Alternative prescription and relation to earlier approaches}
\label{secrel}

In this section we introduce another prescription for calculating the 
quantitities $\chi_i$ and $m$ appearing in \cite{sugita}. It is based on 
the alternative Floquet decomposition
\be
M(t) = \widetilde{M}_{{\rm per}}(t)\, \widetilde{M}_{{\rm av}}(t) 
\equiv \widetilde{M}_{{\rm per}}(t)\, \exp(t\widetilde{K})\,,
\ee
specified by a constant matrix $\widetilde{K}$ such that
\be
\widetilde{K} \xi_i^{\pm} = \pm i \frac{\widetilde{\chi}_i}{T}\, \xi_i^{\pm}\,.
\label{alterk}
\ee
This actually represents another convention for the choice of the stability 
angle. The formula for the Maslov index is then modified to
\be
\sigma_r = \sum_{i=1}^{n_{{\rm ell}}} 
           \left(1 + 2 \left[ \mathrm{sign}(s_i) 
           \frac{r \widetilde{\chi}_i}{2 \pi} \right] \right) 
         + r n_{{\rm ih}} + 2\, \widetilde{m} r\,.            \label{main1} 
\ee
If $s_i>0$ for all $i$, we have $\widetilde{K}=K$ and $\widetilde{\chi}_i 
=\chi_i$, as well as $\widetilde{m}= m$. Then, there is no difference 
between (\ref{main1}) and (\ref{main}). If $s_i<0$ for some $i$, we can 
make the transformation
\be
2 \left[ - \frac{r \tilde{\chi}_i}{2 \pi} \right] 
= -2 r + 2 \left[ \frac{r (2 \pi - \tilde{\chi}_i )}{2 \pi} \right] 
= -2 r + 2 \left[ \frac{r \chi_i }{2 \pi} \right] .           \label{negs}
\ee
Correspondingly, the winding number $m$ of $M_{{\rm per}}(t)$ changes to 
$\widetilde{m}$, which is the winding number of $\widetilde{M}_{{\rm per}}(t)$,
such that
\be
2 \widetilde{m} r = 2 m r +2r\,.                              \label{negm}
\ee
Summing up (\ref{negs}) and (\ref{negm}), we see that the $\sigma_r$ in both 
(\ref{main1}) and (\ref{main}) coincide. Thus, the equivalence of both 
representations is established. We also note that, in general, the difference 
$(\widetilde{m} - m)$ equals to the winding number of $e^{t(K-\widetilde{K})}$,
which is the number of elliptic blocks of $K$ (or $\widetilde{K}$) with 
negative values of $s_i$.

The sign of $s_i$ may change from positive to negative (or vice versa) 
away from bifurcation or symmetry restoring points. As a consequence, 
$\widetilde{m}$ changes its value by $+1$ or $-1$, but such as to conserve 
the total Maslov index. In two-dimensional systems, we have found this to 
happen when the stability discriminant $2-{\rm det}(M_{\rm red}-I_2)={\rm tr}
M_{\rm red}$ crosses or touches the line tr$M_{\rm red}=-2$.
The prescription for $\widetilde{K}$ based on (\ref{alterk}) is not relevant 
from the point of view of a canonically invariant formulation, but such a 
representation often appears to be more convenient in numerical computations. 
It reveals itself useful, in fact, to establish some relations to the approach 
of Creagh, Robbins and Littlejohn \cite{crliro} (see also \cite{wnrobb} and
Appendix D in \cite{book}). As mentioned already above, these authors have
written the Maslov index as a sum \eq{munu} of two contributions, which for
stable orbits must be calculated separately. For stable orbits, $\mu$ and 
$\nu$ are invariants in the sense that they do not depend on the starting
point of the orbit, but they may change their values away from bifurcations 
or symmetry restoring points, exactly as it happens for $\widetilde{m}$. We
note that for a two-dimensional system, $\nu$ is given \cite{crliro} by the 
upper right element $b$ of the stability matrix
\be
M_{red} = \left( \begin{array}{cc}
          a & b \\ c & d   \end{array} \right)
\ee
and can be calculated as
\begin{equation}
b = \frac{\partial r_\perp(T)}{\partial p_\perp(0)}\,,         \label{b}
\end{equation}
where $p_\perp(t)$ and $r_\perp(t)$ are the momentum and coordinate,
respectively, transverse to the orbit. This relation actually 
demonstrates the necessity of knowing the orbit's intrinsic coordinate
system for the calculation of $b$ and hence of $\nu$.

We conclude this section by a number of rules for the Maslov indices
for two-dimensional Hamiltonian systems that are useful for the
classification of periodic orbits, in particular in connection with
complicated bifurcation scenarios such as will be discussed in our 
applications in section \ref{secapp}. The rule 1 is rigorous and follows 
directly from the formulae \eq{main} and \eq{main1}, whereas the rules 
2 -- 4 are empirical, being based on numerical experience. They will be 
illustrated in the examples given in the next section.

\begin{itemize}

\item[1.] In two-dimensional systems, the Maslov index $\sigma_r$ is always
even for hyperbolically unstable orbits and odd for stable and 
inverse-hyperbolically unstable orbits.

\item[2.] For stable orbits, sign$(s)=\,$ sign$(b)$. This means that 
          ${\widetilde \sigma}_{\rm av}$ is always negative when $\nu=0$ 
          and positive when $\nu=1$.

\item[3.] The values of sign$(s)$ of two stable orbits involved in the same 
          bifurcation are identical.

\item[4.] All orbits involved in a bifurcation locally have the same winding
          numbers $\widetilde m$. 

\end{itemize}

These rules are consistent with the fact that Maslov indices of isolated 
orbits can only change at bifurcations and in symmetry-restoring limits. 
The changes at bifurcations are given in the papers of Sieber and Schomerus 
\cite{ssun} and were found to be correctly reproduced by the present 
method in all cases.

\section{Numerical applications} 
\label{secapp}

In this section, we shall apply our method to some systems with mixed
classical dynamics. We first discuss two textbook systems, the
homogeneous two-dimensional quartic oscillator and the famous
H\'enon-Heiles system. They both have Hamiltonians of the form
$H=p^2\!/2 + V(q)$ and have been investigated numerous times in the
framework of periodic orbit theory using the previous methods 
\cite{wint} and \cite{crliro} for the calculation of the Maslov indices. 
In the next two examples we shall study systems with spin degrees of 
freedom, for which a simple separation into kinetic and potential energy 
is not possible and the previous methods are not straightforwardly applied.

\subsection{The quartic oscillator}
\label{secq4}

As a typical system which exhibits the transition from integrable
regular to almost completely chaotic motion, we study the quartic 
oscillator Hamiltonian
\be
H =\frac12\, (p_x^2 + p_y^2)+\frac14\, (x^4+y^4)
  +\frac{\alpha}{2}\, x^2 y^2\,.
\label{hamq4}
\ee
It is homogeneous in coordinates and momenta, so that the energy can be 
scaled away. The chaoticity parameter is $\alpha$. For $\alpha=0$, 1
and 3, the system is integrable and in the limits $\alpha\rightarrow -1$
and $\alpha\rightarrow\infty$ it becomes nearly chaotic \cite{daru}. The 
stability of the linear orbit running along either of the axes, which we 
here denote by A, is known analytically \cite{yosh}. The trace of its 
stability matrix is given by $\trM = 4\,\cos(\pi\sqrt{1+8\alpha}/2) 
+ 2$. Isochronous bifurcations of the primitive A orbit, which are of
pitchfork type, occur when $\trM$ takes the value +2 which occurs at the 
values $\alpha_n = \frac12\,n\,(n+1)$ with $n=0,3,4,5,\dots$ Period-doubling 
bifurcations of island-chain type occur at the values $\alpha_p=2p\,(p+1)+3/8$ 
with $p=0,1,2,\dots$ These bifurcations of the A orbit and the analytical
properties of the period-one and period-two orbits created at the 
bifurcations have been discussed in \cite{lamp} and \cite{bfmm}. As an 
example for our present evaluation of the winding number $m$, we shown in 
the left part of \fig{phiq4} the function $\varphi(t)$ defined in \eq{phit},
obtained here for the stable orbit L$_7$ born at the bifurcation of the A 
orbit at $\alpha=15$. It yields $m=3$. Since $s>0$ in this case, $\varphi(t)$ 
is identical to $\widetilde{\varphi}(t)$. The Maslov index becomes 
$\sigma_1=7$, in agreement with the value obtained in \cite{lamp} using the 
formalism of \cite{crliro}. In the right part of \fig{phiq4} we show the 
shape of the L$_7$ orbit in the $(x,y)$ plane. 

\Figurebb{phiq4}{-60}{20}{795}{330}{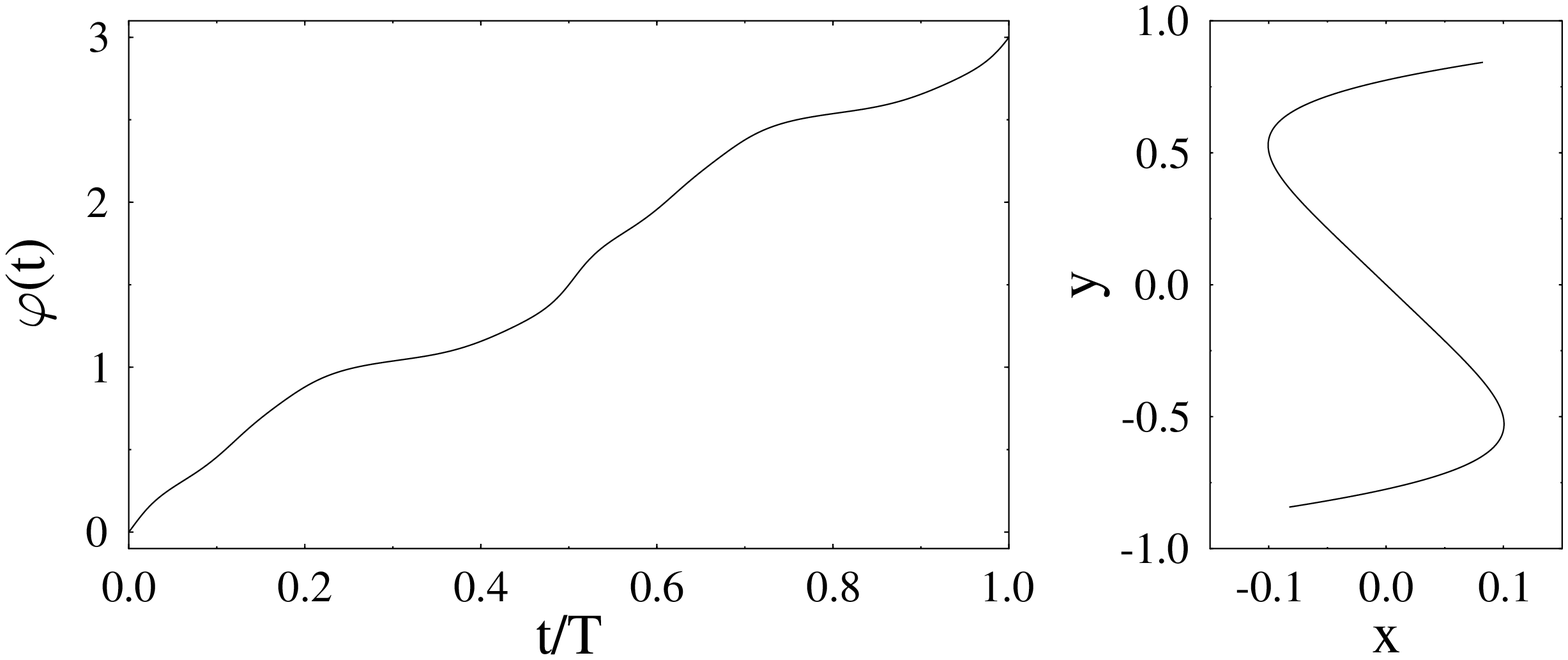}{5}{16.6}{
Properties of the the L$_7$ orbit in the quartic oscillator 
Hamiltonian \eq{hamq4} at $\alpha=16.8$. {\it Left panel:} 
phase function $\varphi(t)$ \eq{phit} giving $m=3$; {\it right panel:} 
shape of the L$_7$ orbit in the $(x,y)$ plane.
}

In \cite{lakh}, the scaling behaviour of the fix points corresponding 
to period-four orbits created at bifurcations of the fourth repetition
of the A orbit (denoted here by A$^4$) have been discussed. To illustrate 
this scenario, we show in \fig{q4trm4} the stability discriminant $\trM$ 
of the period-four orbits involved in an island-chain bifurcation of the 
A$^4$ orbit, which occurs at $\alpha=5.4305556$, and in the succeeding 
pitchfork bifurcations of the P$_{21}$ and P$'_{21}$ orbits. The 
subscripts of the orbit names indicate their Maslov indices as obtained 
using the formulae of \cite{crliro}. (Not shown are the orbits created at 
the bifurcations of A$^4$ occurring at $\alpha=4.375$ and at $\alpha=6$.)

\Figurebb{q4trm4}{25}{40}{852}{495}{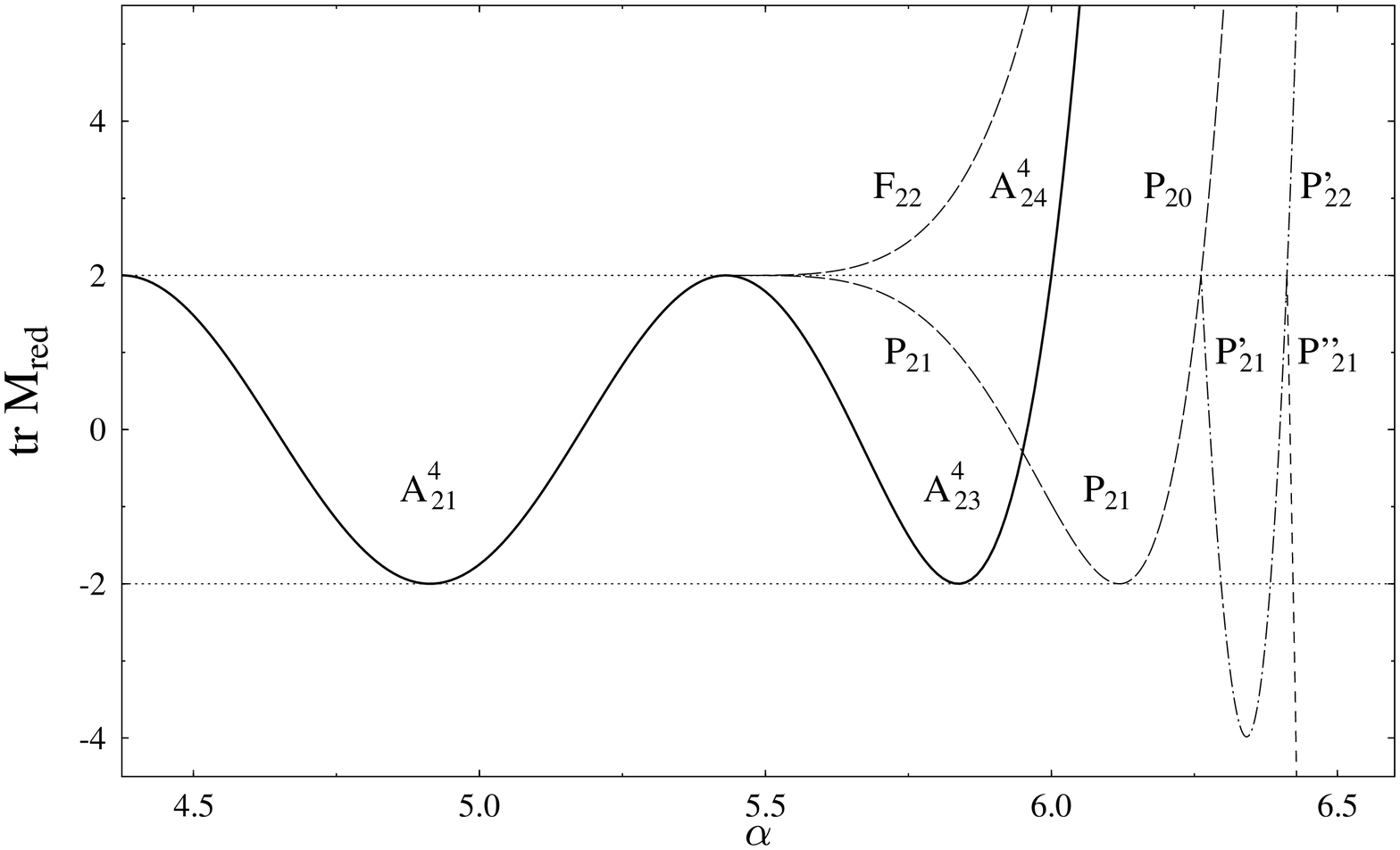}{9}{16.6}{
Stability discriminant $\trM$ of the period-four orbits in the 
quartic oscillator \eq{hamq4} involved in an island-chain bifurcation 
of the A$^4_{21}$ orbit occurring at $\alpha=5.4305556$ and in the 
succeeding pitchfork bifurcations of the P$_{21}$ and P$'_{21}$ orbits.
}

\Figurebb{q4pq6}{-20}{10}{810}{500}{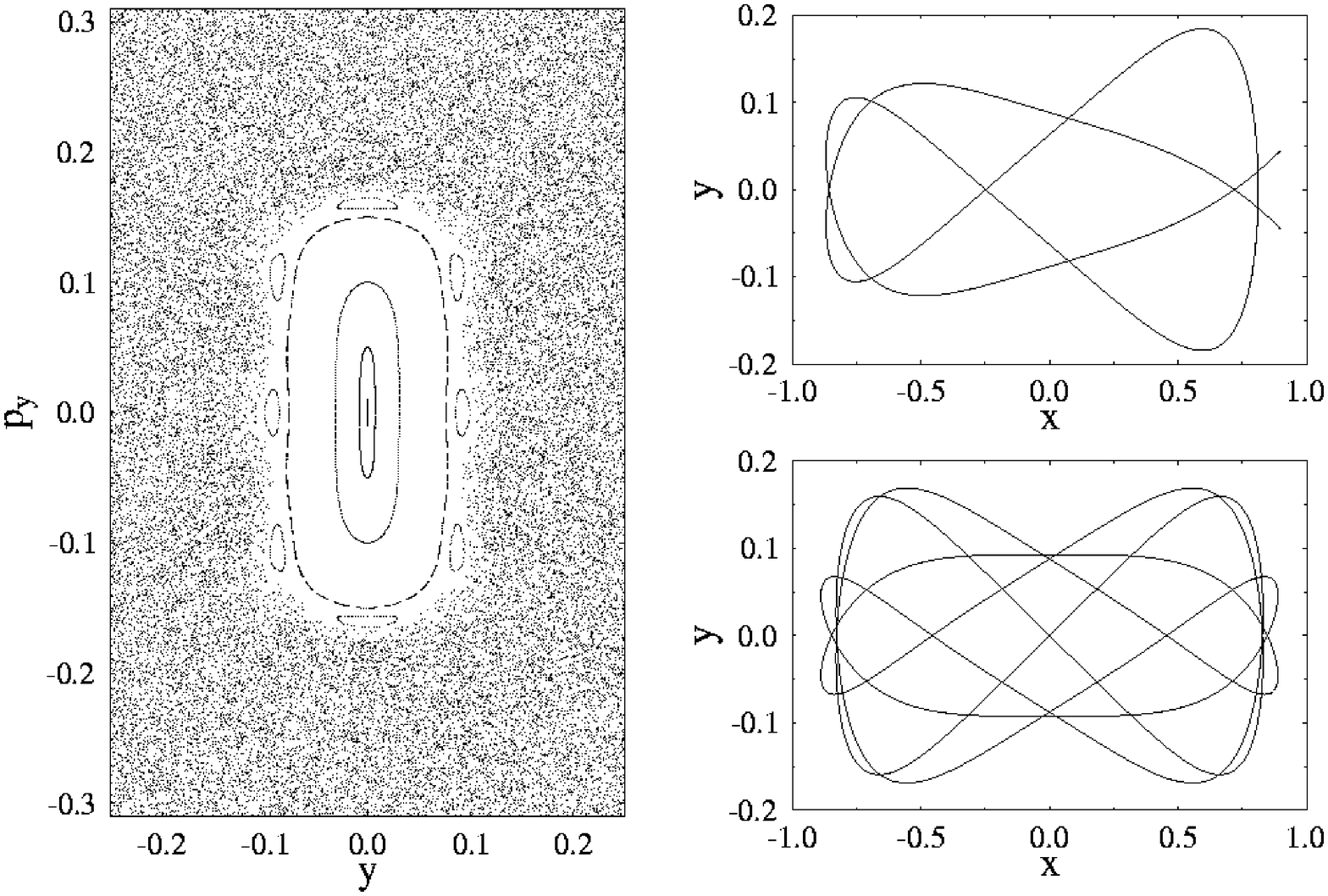}{10.2}{16.6}{
Poincar\'e surface of section $(y,p_y)$, shown in the left panel,
taken at $x=0$ in the quartic oscillator \eq{hamq4} at $\alpha$=6.0. 
The fix point in the central stability island corresponds to the A 
orbit, and the chain of eight unstable and stable fix points correspond 
to the two pairs of period-four orbits F$_{22}$ and P$_{21}$ with shapes 
shown in the upper and lower right panels, respectively. 
}

In \fig{q4pq6} we show on the left side a Poincar\'e surface of section
$(y,p_y)$, taken for $x=0$ at $\alpha=6.0$. The fix point in the central 
stability island corresponds to the A orbit and its repetitions. The KAM
chain of eight unstable and stable fix points, which form the boundary
of the stability island towards the chaotic sea, correspond to the two 
pairs of period-four orbits F$_{22}$ and P$_{21}$ whose $(x,y)$ shapes 
are shown in the upper and lower right panels of \fig{q4pq6}, 
respectively. These are the fix points whose scaling with $\alpha$ was
studied in \cite{lakh}.

Table \ref{q4bif22} shows the stabilities, Maslov indices and their 
ingredients of the above period-four orbits in the quartic oscillator. 
We give the intervals $(\alpha_{min},\alpha_{max})$ in which the orbits 
are stable (ell) with a fixed sign of $s$, 
hyperbolically (hyp) unstable, or inverse-hyperbolically (i-hyp) 
unstable. Values of $(\alpha_{min}$ and $\alpha_{max})$ marked by an 
asterisk (*) denote bifurcation points for the orbits listed in the 
corresponding rows. Note that the values of $\mu$ and $\nu$ are not 
unique for unstable orbits; they may depend on the starting point along 
the orbit chosen for their calculation, but such that $\sigma=\mu+\nu$ 
is invariant \cite{crliro}. For stable orbits, they are constant in each 
of the given regions and related to $2 \widetilde{m}$ and 
$\widetilde{\sigma}_{\mathrm{av}}$ as discussed in section \ref{secrel}.
Our results show the consistent agreement between the definitions
of the Maslov index by Creagh et al \cite{crliro} and by Sugita 
\cite{sugita}. They also illustrate the empirical rules given at the
end of section \ref{secrel}.

\Table{q4bif22}{15}{
\begin{tabular}{|l|l|l|c|c|c|c|r|c|c|c|} \hline
orbit & $\alpha_{min}$  &  $\alpha_{max}$ & stab & $m$ & $\sigma_{\mathrm{av}}$
 & 
$\widetilde{m}$ & $\widetilde{\sigma}_{\mathrm{av}}$ &  $\sigma$ & $\mu$ & 
$\nu$ \\ \hline\hline
A$^4_{21}$ & 4.375* & 4.913    & ell   & 10  & 1 & 10  & +1   & 21 & 20 & 1 \\
 \hline
A$^4_{21}$ & 4.913  & 5.431*   & ell   & 10  & 1 & 11  & $-1$ & 21 & 21 & 0 \\
 \hline
A$^4_{23}$ & 5.431* & 5.837    & ell   & 11  & 1 & 11  & +1   & 23 & 22 & 1 \\
 \hline
A$^4_{23}$ & 5.837  & 6.0*     & ell   & 11  & 1 & 12  & $-1$ & 23 & 23 & 0 \\
 \hline
A$^4_{24}$ & 6.0*   & 10.0*    & hyp   & 12  & 0 & 12  & 0    & 24 & 23/24 & 1/
0\\ \hline
F$_{22}$   & 5.431* & $\infty$ & hyp   & 11  & 0 & 11  & 0    & 22 & 21/22 & 1/
0\\ \hline
P$_{21}$   & 5.431* & 6.118    & ell   & 10  & 1 & 11  & $-1$ & 21 & 21 & 0 \\
 \hline
P$_{21}$   & 6.118  & 6.262    & ell   & 10  & 1 & 10  & +1   & 21 & 20 & 1 \\
 \hline
P$_{20}$   & 6.262  & $\infty$ & hyp   & 10  & 0 & 10  & 0    & 20 & 19/20 & 1/
0\\ \hline
P$'_{21}$  & 6.262  & 6.341    & ell   & 10  & 1 & 10  & +1   & 21 & 20 & 1 \\
 \hline
P$'_{21}$  & 6.341  & 6.383    & i-hyp & 10  & 1 & 10  & +1   & 21 & 20/21 & 1/
0\\ \hline
P$'_{21}$  & 6.383  & 6.412*   & ell   & 10  & 1 & 11  & $-1$ & 21 & 21 & 0 \\
 \hline
P$'_{22}$  & 6.412* & $\infty$ & hyp   & 11  & 0 & 11  & 0    & 22 & 21/22 & 1/
0\\ \hline
P$''_{21}$ & 6.412* & 6.422    & ell   & 10  & 1 & 11  & $-1$ & 21 & 21 & 0 \\
 \hline
P$''_{21}$ & 6.422  & $\infty$ & i-hyp & 10  & 1 & 10  & +1   & 21 & 20/21 & 1/
0\\ \hline
\end{tabular}
}{~~Stabilities, Maslov indices and related properties of the orbits in the
quartic oscillator whose stability discriminants $\trM$ are shown in 
\fig{q4trm4}. 'ell', 'hyp' and 'i-hyp' denote elliptic (stable), hyperbolic 
and inverse-hyperbolic (unstable) orbits, respectively. Values of 
$\alpha_{min}$ or $\alpha_{max}$ marked by an asterisk (*) denote bifurcation 
points for the orbits given in the corresponding row.
For unstable orbits, the decomposition of $\sigma=\mu+\nu$ is not unique.
}

\subsection{The H\'enon-Heiles system}
\label{sechh}

Another famous system with mixed classical dynamics is given by the 
H\'enon-Heiles Hamiltonian \cite{hh}
\be
H =\frac12\, (p_x^2 + p_y^2)+\frac12\, (x^2 + y^2) +
   \varepsilon\,(x^2 y -\frac13\, y^3)\,,                     \label{hamhh}
\ee
where $\epsilon$ regulates the chaoticity of the system. The potential 
in \eq{hamhh} has three saddle points at the energy 
$E^*=1/6\,\varepsilon^2$, over which a particle can escape if $E>E^*$. 
The classical dynamics depend only on the scaled energy 
$e=E/E^*=6\,\varepsilon^2E$; in this variable the saddles are at $e=1$. 
Along the symmetry lines passing through the saddles, one of them being 
the $y$ axis, there are librating orbits (denoted here again by A) whose 
stability oscillates infinitely many times as the energy approaches the 
critical value $e=1$, giving rise to an infinite cascade of isochronous 
pitchfork bifurcations. The scaled bifurcation energies $e_n$ ($n=1,2,
\dots,\infty$) form a sequence that cumulates at $e_\infty=1$ in a 
Feigenbaum-like fashion; the orbits born at the bifurcations exhibit 
self-similarity with analytically known scaling constants \cite{mbgu}. 

At each successive bifurcation $e_n$, the A orbit increases its Maslov 
index by one unit. The orbits born at the bifurcations are alternatingly 
stable rotations R$_\sigma$ and unstable librations L$_\sigma$; they can be 
uniquely classified by their increasing Maslov indices: R$_5$, L$_6$, R$_7$, 
L$_8$, etc, according to the rules given at the end of section \ref{secrel}
(cf also \cite{lamp,mbgu}). Besides the A orbit, the system possesses a 
curved librating orbit B which is unstable at all energies, and a rotating 
orbit C which is stable up to $e=0.8919$ where it turns inverse 
hyperbolically unstable. Its 

\Table{hhorb}{15}{
\begin{tabular}{|l|l|l|c|c|c|c|r|c|c|c|} \hline
orbit & $e_{min}$  &  $e_{max}$ & stab & $m$ & $\sigma_{\mathrm{av}}$ & 
$\widetilde{m}$ & $\widetilde{\sigma}_{\mathrm{av}}$ & $\sigma$ & $\mu$ & 
$\nu$  \\ \hline\hline
A$_{5}$ & 0.0     & 0.8117   & ell   & 2 & 1 & 2  & +1   & 5 & 4 & 1 \\ \hline
A$_{5}$ & 0.8117  & 0.9152   & i-hyp & 2 & 1 & 2  & +1   & 5 & 4/5 & 1/0 \\ 
\hline
A$_{5}$ & 0.9152  & 0.9693*  & ell   & 2 & 1 & 3  & $-1$ & 5 & 5 & 0 \\ \hline
A$_{6}$ & 0.9693* & 0.9867*  & hyp   & 3 & 0 & 3  & 0    & 6 & 5/6 & 1/0 \\ 
\hline
R$_{5}$ & 0.9693* & 0.9895   & ell   & 2 & 1 & 3  & $-1$ & 5 & 5 & 0 \\ \hline
R$_{5}$ & 0.9895  & $\infty$ & i-hyp & 2 & 1 & 2  & +1   & 5 & 4/5 & 1/0 \\ 
\hline
A$_{7}$ & 0.9867* & 0.9950   & ell   & 3 & 1 & 3  & +1   & 7 & 6 & 1 \\ \hline
L$_{6}$ & 0.9867* & $\infty$ & hyp   & 3 & 0 & 3  & 0    & 6 & 5/6 & 1/0 \\ 
\hline
A$_{7}$ & 0.9950  & 0.9978   & i-hyp & 3 & 1 & 3  & +1   & 7 & 6/7 & 1/0 \\ 
\hline
A$_{7}$ & 0.9978  & 0.9992*  & ell   & 3 & 1 & 4  & $-1$ & 7 & 7 & 0 \\ \hline
A$_{8}$ & 0.9992* & 0.9996*  & hyp   & 4 & 0 & 4  & 0    & 8 & 7/8 & 1/0 \\ 
\hline
R$_{7}$ & 0.9992* & 0.99948  & ell   & 3 & 1 & 4  & $-1$ & 7 & 7 & 0 \\ \hline
R$_{7}$ & 0.99948 & $\infty$ & i-hyp & 3 & 1 & 3  & +1   & 7 & 6/7 & 1/0 \\ 
\hline
B$_4$   & 0.0     & $\infty$ & hyp   & 2 & 0 & 2  & 0    & 4 & 3/4 & 1/0 \\ 
\hline
C$_{3}$ & 0.0     & 0.8921   & ell   & 1 & 1 & 2  & $-1$ & 3 & 3 & 0 \\ \hline
C$_{3}$ & 0.8921  & $\infty$ & i-hyp & 1 & 1 & 1  & +1   & 3 & 2/3 & 1/0 \\ 
\hline
C$^2_7$ & 0.0     & 0.6146   & ell   & 3 & 1 & 4  & $-1$ & 7 & 7 & 0 \\ \hline
C$^2_7$ & 0.6146  & 0.8921*  & ell   & 3 & 1 & 3  & +1   & 7 & 6 & 1 \\ \hline
C$^2_6$ & 0.8921* & $\infty$ & hyp   & 3 & 0 & 3  & 0    & 6 & 5/6 & 1/0 \\ 
\hline
D$_7$   & 0.8921* & 1.013    & ell   & 3 & 1 & 3  & +1   & 7 & 6 & 1 \\ \hline
D$_7$   & 1.013   & 1.180*   & ell   & 3 & 1 & 4  & $-1$ & 7 & 7 & 0 \\ \hline
D$_9$   & 1.180*  & 1.2375   & ell   & 4 & 1 & 4  & +1   & 9 & 8 & 1 \\ \hline
D$_9$   & 1.2375  & $\infty$ & i-hyp & 4 & 1 & 4  & +1   & 9 & 8/9 & 1/0 \\ 
\hline
\end{tabular}
}{Shortest period-one and period-two orbits in the H\'enon-Heiles
system, their stabilities, Maslov indices and related properties.
Notation as in \tab{q4bif22}. $e=E/E^*$ is the scaled energy;
its values denoted by asterisks (*) are bifurcation energies $e_n$.
}

\vspace*{-0.5cm}

\noindent
second
repetition bifurcates at this energy, giving birth to an orbit D that 
stays stable up to $e=1.2375$ where it becomes inverse hyperbolically 
unstable. We have calculated the Maslov index of all these orbits using 
the formulae given above and verified that they agree with the values 
obtained in \cite{hhun,hhpo,mbgu} using the method of \cite{crliro} and 
in \cite{lamp} using the method of \cite{wint}. The results are given 
in \tab{hhorb}, again in energy intervals of constant 
$\widetilde{\sigma}_{\mathrm{av}}$, $\mu$ and $\nu$.

Although tangent bifurcations are known to occur generically in chaotic
and mixed-dyn\-amical systems, to our knowledge no such bifurcation has been 
reported so far in the H\'enon-Heiles system. In \cite{lamp,mbgu} we have 
wrongly surmised that all its periodic orbits existing below the barrier
energy $e=1$ are derivatives of the generic orbits A, B, C (and their
repetitions) through their bifurcations. This was not correct, as we can 
demonstrate in the following two figures. Here we present a sequence of 
``spider''-like orbits that are born out of a tangent bifurcation occurring
at the scaled energy $e=0.988249$. Their stability discriminants
tr$M_{\rm red}$ are shown in \fig{spidtrm}, and their six genuine
shapes in the $(y,x)$ plane in \fig{spidxy}. The generic pair of {\bf a} 
orbits, born with Maslov indices 22 and 23, keeps its shape through three
successive pitchfork bifurcations at which the orbits {\bf c}, {\bf b} and
{\bf a'} are born; these five orbits remain hyperbolically unstable
at all energies $e\simg 1.016$. The latter two bifurcate again, 
giving birth

\Figurebb{spidtrm}{-10}{45}{739}{487}{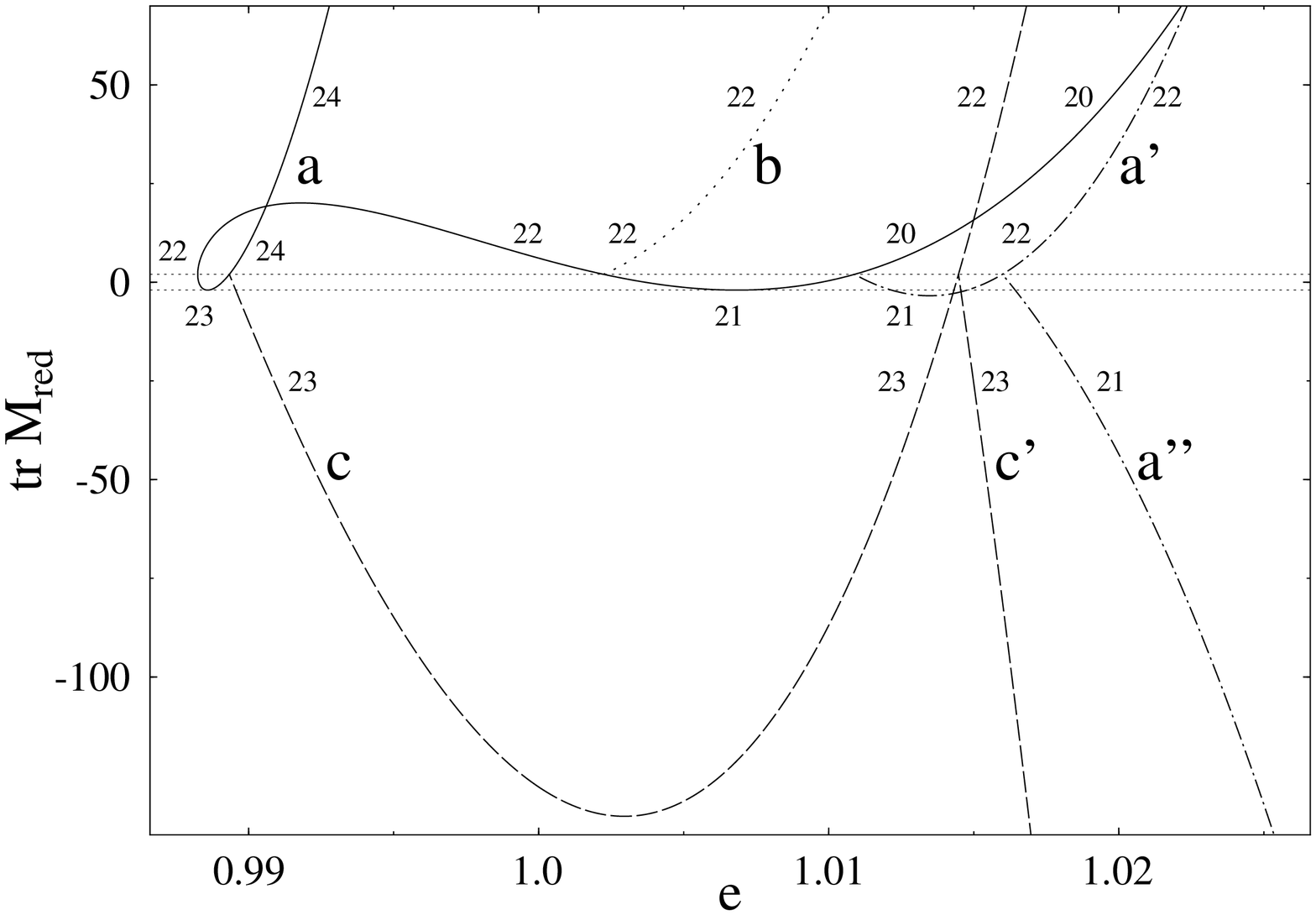}{8.9}{16.6}{
Stability discriminants tr$M_{\rm red}$ of the sequence of ``spider'' 
orbits in the H\'enon-Heiles potential born out of a tangent bifurcation 
at the scaled energy $e=0.988249$. The numbers give their Maslov indices 
$\sigma$ in the respective stability intervals; the letters refer to the 
six orbit types whose shapes are shown in \fig{spidxy}; the horizontal 
dotted lines correspond to tr$M_{\rm red}=\pm 2$.
}

\noindent
to the orbits {\bf c'} 
and {\bf a''} which remain inverse-hyperbolically unstable at all 
energies $e\simg 1.016$. Note that the three orbits {\bf a, b, c} are 
reflection-symmetric around the symmetry axes (shown by the dotted lines 
in \fig{spidxy}) containing the A orbits, whereas the others are not. The 
Maslov indices, which fulfill again the rules of section \ref{secrel},
have been obtained with our present method. We found, in fact, that the
earlier methods of \cite{wint} and \cite{crliro} could not be applied
safely here: the use of the intrinsic coordinate systems of these complicated 
orbits is numerically not always stable enough to yield unique results. This
actually demonstrates an advantage of the present method which works
reliably for not too unstable orbits ($|$tr$M_{\rm red}|\siml 40$).

\Figurebb{spidxy}{-120}{20}{795}{550}{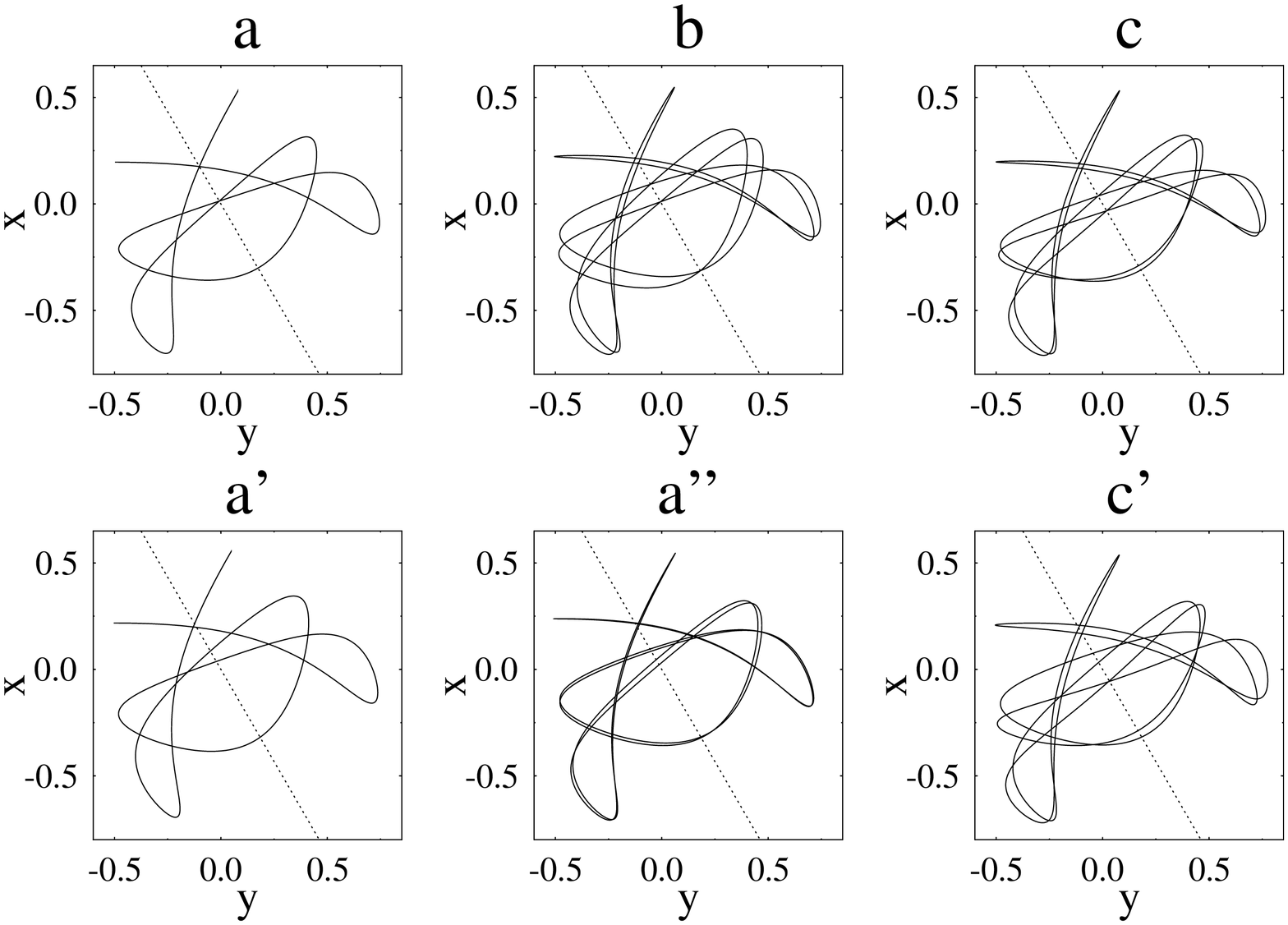}{8.}{16.6}{
Shapes of the ``spider'' sequence of periodic orbits in the H\'enon-Heiles 
potential.
}

\subsection{A spin-boson system}
\label{secdicke}

We consider a spin-boson system defined by the quantum Hamiltonian 
\be
\hat{H} = \omega_b\, \hat{a}^{\dag} \hat{a}+ \omega_s\, \hat{s}_3 + \kappa\,
(\hat{a}^{\dag} +\hat{a}) (\hat{s}_{+}+ \hat{s}_{-})\,, \qquad
\qquad {\hat s}_\pm = \hat{s}_1\pm i\hat{s}_2\,,
\label{rabi}
\ee
where $\hat{s}_{\alpha} = \frac12 \sigma_{\alpha} \,\, (\alpha=1,2,3)$ 
are the usual spin operators for $s=\frac12$ particles ($\hbar=1$).
This model has a broad range of applications in atomic, molecular and
solid-state physics and in quantum optics. In the different fields the
Hamiltonian (\ref{rabi}) bears different names, among them ``Rabi
Hamiltonian'' and ``molecular polaron model'' (see \cite{grah} for
a review and further references). 

In order to treat this system (semi-)classically, we have to define
a phase-space symbol for the Hamiltonian (\ref{rabi}). To this
purpose we introduce the bosonic operators $\hat{a}=(\hat{Q}_b+ 
i\hat{P}_b)/\sqrt{2}$, $\hat{a}^{\dag}=(\hat{Q}_b-i\hat{P}_b)/\sqrt{2}$
and take their Wigner transforms to define the canonical bosonic
variables $(Q_b,P_b)$. For the spin variables $n_{\alpha}$ we use the spin 
coherent-state symbols of the spin operators $\hat{s}_{\alpha}$, divided 
by the value of spin $s$ (see eg \cite{plet}). This leads to the following
symbol of the Hamiltonian (\ref{rabi})
\be
H = \frac{\omega_b}{2}\, (P_b^2 + Q_b^2 -1) + \frac{\omega_s}{2}\, n_3 +
    \sqrt{2}\kappa \,Q_b \,n_1\,.  
\label{class}
\ee
The classical equations of motion read
\bea
\dot{Q}_b &=& \omega_b P_b\,, \qquad \dot{P}_b \;=\; -\omega_b Q_b-\sqrt{2}
\kappa n_1\,, \label{boseq} \\ 
\dot{n}_1 &=& -\omega_s n_2\,,\quad\;\;\dot{n}_2
              \;=\;\omega_s n_1-\kappa Q_b n_3/\sqrt{2}\,, \quad\;\; 
    \dot{n}_3 \;=\; \kappa Q_b n_2 /\sqrt{2}\,, 
\label{spieq}
\eea
with the constraint $n_1^2 +n_2^2 +n_3^2 =1$.

We can now introduce the Darboux coordinates $(Q_s,P_s)$ by making a 
stereographic projection from the North Pole of the unit ${\bf n}$-sphere onto 
the complex plane and then contracting the plane to a disc with radius
$\sqrt{2}$: 
\be
n_1 = P_s \sqrt{2 - P_s^2 - Q_s^2}\,, \qquad   
n_2 = Q_s \sqrt{2 - P_s^2 - Q_s^2}\,, \qquad
n_3 = P_s^2 + Q_s^2 -1\,.                                  
\label{dar1}
\ee
Under this mapping, the North Pole is projected onto the boundary of the
disc $P_s^2 + Q_s^2 =2$, and the South Pole is projected into the centre
of the disc. The Hamiltonian then has the form
\be
H = \frac{\omega_b}{2}\, (P_b^2 + Q_b^2 - 1) 
  + \frac{\omega_s}{2}\, (P_s^2 + Q_s^2 - 1)
  + \kappa\,Q_b P_s \sqrt{2 (2 - P_s^2 - Q_s^2 )}\,.  
\label{sbclass}
\ee
This is a two-dimensional harmonic oscillator, perturbed by the
nonlinear term proportional to $\kappa$.
The representation \eq{dar1} is convenient because the equations of motion 
(\ref{boseq}), (\ref{spieq}) for both boson and spin variables can
now be written in a canonical Hamiltonian form: 
\be
\dot{Q}_a = \frac{\partial H}{\partial P_a}, \qquad \dot{P}_a =
-\frac{\partial H}{\partial Q_a}, \qquad\qquad  (a=b,s)
\label{eqpl}
\ee
and the equation for the matrizant $M(t)$ is easily found.

However, as soon as we cross the North Pole on the ${\bf n}$-sphere, the
equations (\ref{eqpl}) become singular, and one has to switch to the
alternative representation
\be
n_1 = P_s \sqrt{2 - P_s^2 - Q_s^2}\,, \qquad   
n_2 = -Q_s \sqrt{2 - P_s^2 - Q_s^2}\,, \qquad
n_3 = 1 - P_s^2 - Q_s^2\, ,
\label{dar2}
\ee
which corresponds to the projection from the South Pole.

The ($Q_b$, $P_b$) sections of the periodic orbits in this system are
similar to those of the orbits in the unperturbed harmonic oscillator 
($\kappa=0$), while the spin components $n_{\alpha}$ on the sphere --
or, correspondingly, on the ($Q_s$, $P_s$) disc -- evolve substantially 
with increasing coupling constant $\kappa$. For small $\kappa$, the two 
periodic orbits R$_3$ and R$_5$ originate from the South and the North 
Pole, respectively (\fig{spin3r5}). In the ($Q_b,Q_s$) space they are 
simple rotations with Maslov indices 3 and 5, respectively (\fig{orbr3r5}), 
becoming more and more distorted with increasing $\kappa$. At larger values 
of $\kappa$ they undergo pitchfork bifurcations (see \fig{sbstab}), giving 
birth to the orbits P$_3$ and Q$_5$, respectively. The ($Q_b,Q_s$) shapes 
of the four orbits R$_2$, R$_6$, P$_3$ and Q$_5$ at $\kappa=0.27$ are shown 
in \fig{dishap}. (The subscripts in R$_\sigma$ and Q$_\sigma$ denote again 
the Maslov indices of the respective orbits.)

We note that in the limit $\kappa \to 0$ the Maslov index of the orbit 
R$_3$ coincides with the Maslov index ($\sigma=3$) of the shortest isolated 
orbit in the unperturbed harmonic oscillator (\ref{anisharm}) with the 
frequency ratio $\omega_s:\omega_b=0.6$. The orbit R$_5$ is ill-defined in 
the representation (\ref{dar1}) in this limit, and we have to switch to 
(\ref{dar2}) instead. For the Hamiltonian (\ref{class}), this is equivalent 
to changing $\omega_s \to - \omega_s$. Then, the formula (\ref{anisharm}) 
yields $\sigma=1$ which is equal to 5 (mod 4). The difference can be 
associated with the Maslov index of the matrix which transforms the matrizant 
in the representation (\ref{dar1}) to the matrizant in the representation 
(\ref{dar2}), even though this matrix is not defined at $\kappa=0$. 

\vspace*{0.5cm}
  
\hspace{-0.4cm}
\begin{minipage}{7cm}
\Figurebb{orbr3r5}{-40}{3}{258}{188}{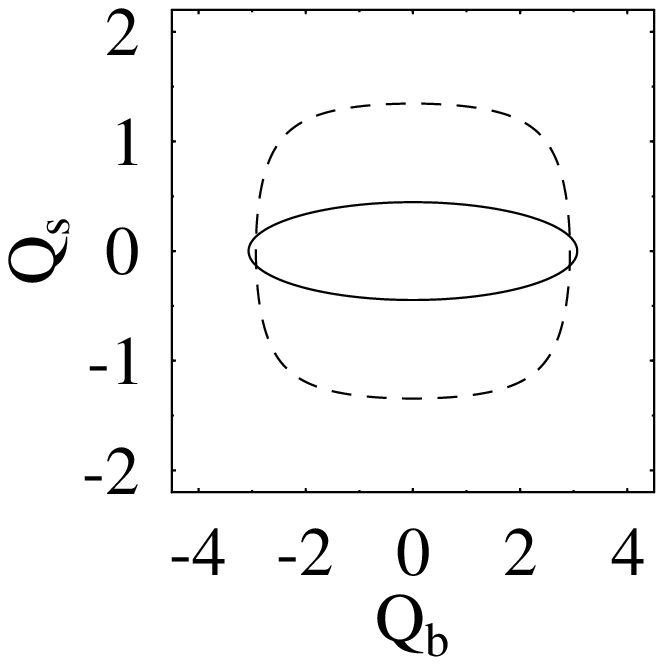}{4.2}{7.5}{
($Q_b,Q_s$) shapes of orbits R$_3$ (solid line) and R$_5$ (dashed line)
in the spin-boson system \eq{sbclass} at $\kappa=0.05$. Other parameters 
as in \fig{dishap}.}
\end{minipage}\hspace{1.2cm}
\begin{minipage}{7cm}
\Figurebb{spin3r5}{50}{3}{258}{188}{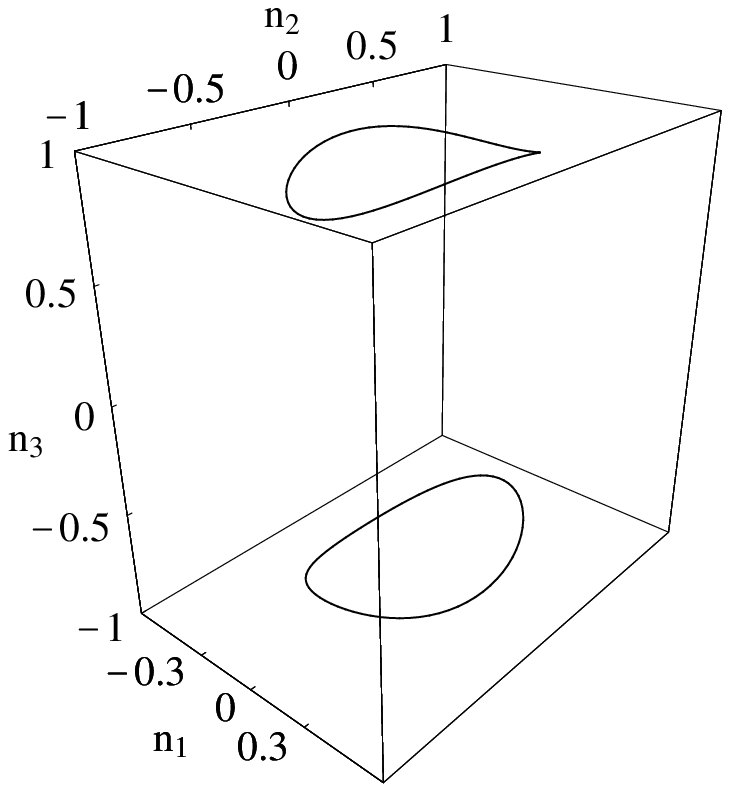}{4.6}{7.5}{
Spin components $n_{\alpha}$ of the orbits R$_3$ and R$_5$ in the spin-boson 
system  \eq{sbclass} at $\kappa=0.05$. Other parameters as in \fig{dishap}.}
\end{minipage}

\Figurebb{dishap}{-10}{35}{850}{260}{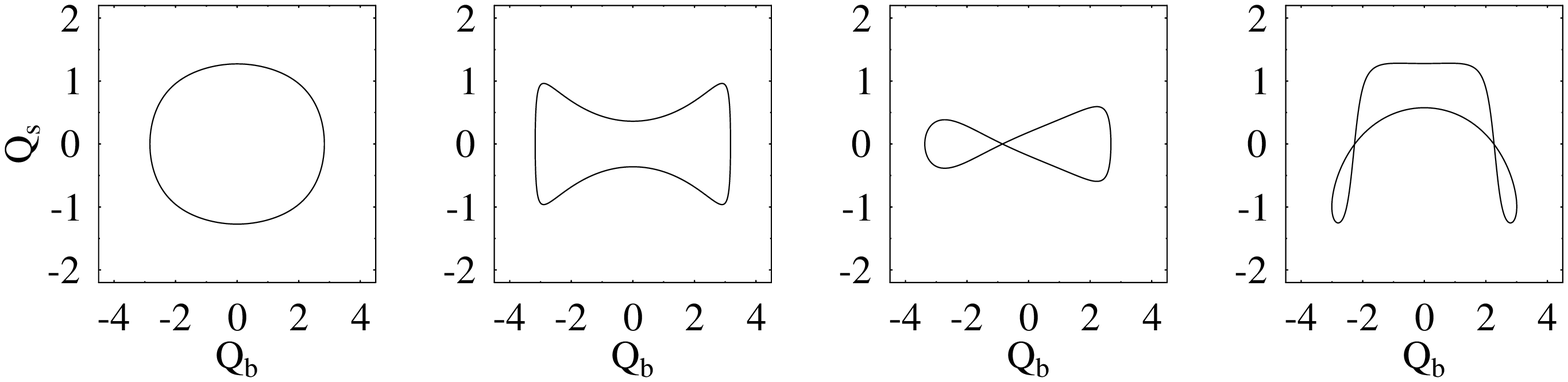}{4.5}{16.6}{
Shapes $Q_s$ versus $Q_b$ of the four orbits R$_2$, R$_6$, P$_3$ and
Q$_5$ (from left to right) in the spin-boson model \eq{sbclass}
at $\kappa=0.27$. Other parameter values: $\omega_b =1.0$, $\omega_s=0.6$, 
$E=4.0$.
}

\Figurebb{sbstab}{-10}{40}{850}{490}{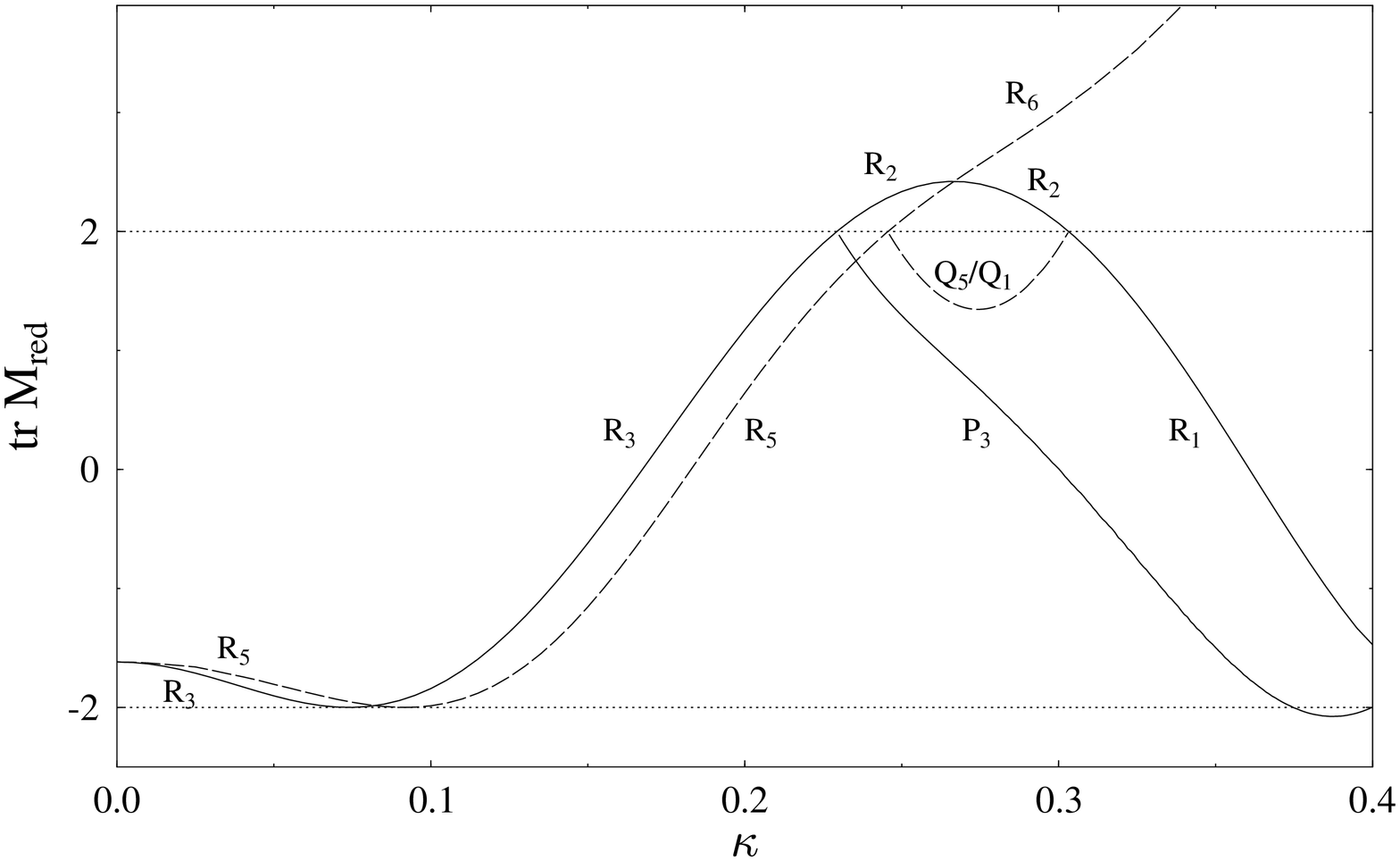}{8.5}{16.6}{
Stability discriminants of the shortest periodic orbits in the 
spin-boson Hamiltonian \eq{sbclass} with versus $\kappa$. Other 
parameters as in \fig{dishap}. The subscripts denote the Maslov indices.
}

The bifurcation scenario is shown in \fig{sbstab}, where we plot the 
discriminants tr$M_{\rm red}$ of these orbits versus the parameter $\kappa$. 
The orbits R$_3$ and R$_5$ touch the line $\trM=-2$ in the stability 
diagram due to the presence of the discrete reflection symmetry $Q_b \to 
-Q_b$, $P_s \to -P_s$, $t \to -t$. The new orbits P$_3$ and Q$_5$ born at 
their bifurcations have more complicated self-crossing rotational shapes 
in the ($Q_b,Q_s$) space with a lower discrete symmetry than that of R$_3$ 
and R$_5$ (see \fig{dishap}). The shape of R$_3$ does not change 
qualitatively after its successive bifurcations, when it becomes R$_2$ and 
R$_1$. The same holds for R$_5$ which becomes R$_6$. 

The orbit Q$_5$/Q$_1$ is interesting in the sense that for $\kappa =0.27$ 
it has the Maslov index $\sigma=5$, while at $\kappa =0.30$ its 
Maslov index is $\sigma=1$. The sign of $s$ is negative in the entire 
interval of existence of this orbit; the change in the Maslov index 
$\widetilde{m}$ is due to a drop of the winding number from 3 to 1 near 
$\kappa \simeq 0.275$. This sudden change of the Maslov index by four 
units, without bifurcation, can be accounted for by a touching 
of the North Pole near $\kappa \simeq 0.275$. It reflects the 
singularity of the representation (\ref{dar1}) and is not felt in the 
semiclassical trace formula \eq{tfiso} where the Maslov enters only
modulo multiples of four.
  
\hspace{-0.4cm}
\begin{minipage}{7cm}
\Figurebb{dio1}{50}{3}{258}{188}{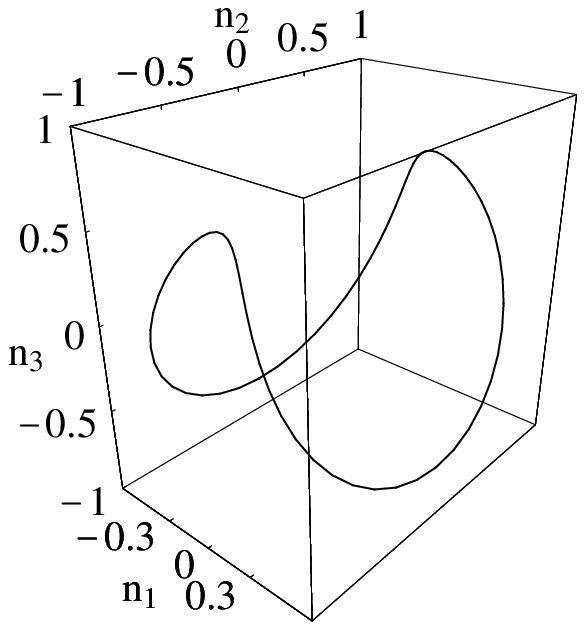}{5.2}{7.5}{
Spin components $n_{\alpha}$ of the orbit R$_2$ in the spin-boson system 
\eq{sbclass} at $\kappa=0.27$. Other parameters as in \fig{dishap}.}
\end{minipage}\hspace{1.2cm}
\begin{minipage}{7cm}
\Figurebb{dio2}{50}{3}{258}{188}{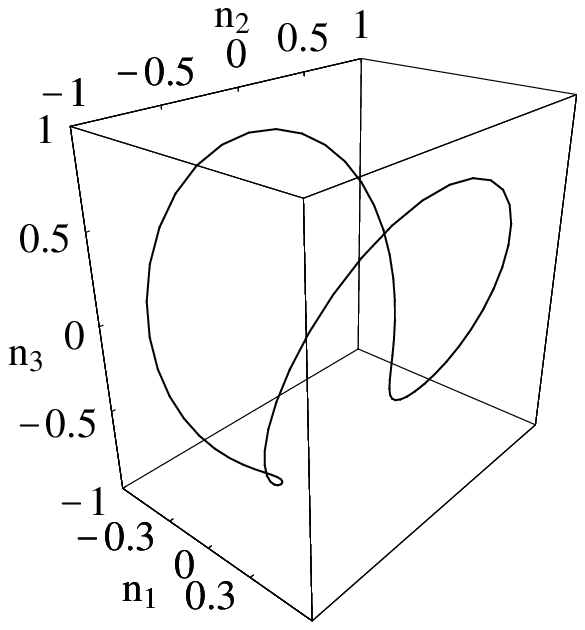}{5.2}{7.5}{
Spin components $n_{\alpha}$ of the orbit R$_6$ in the spin-boson system 
\eq{sbclass} at $\kappa=0.27$. Other parameters as in \fig{dishap}.}
\end{minipage}

The spin components $n_{\alpha}$ of the orbits R$_2$ and R$_6$ at 
$\kappa=0.27$ are shown in figures \ref{dio1} and \ref{dio2} and those 
of the orbit Q$_5$/Q$_1$ at $\kappa=0.27$ and 0.30, respectively, in 
figures \ref{dio3} and \ref{dio4}. 

\hspace{-0.4cm}
\begin{minipage}{7cm}
\Figurebb{dio3}{50}{3}{258}{188}{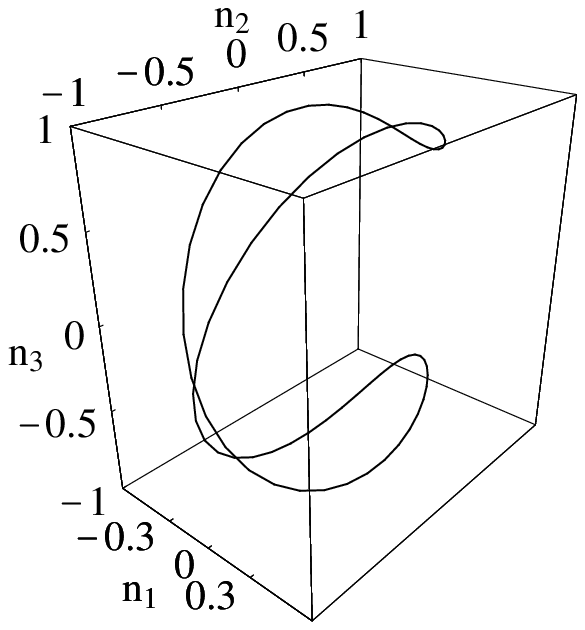}{5.2}{7.5}{
Spin components $n_{\alpha}$ of the orbit Q$_5$ in the spin-boson system 
\eq{sbclass} at $\kappa=0.27$. Other parameters as in \fig{dishap}.}
\end{minipage}\hspace{1.2cm}
\begin{minipage}{7cm}
\Figurebb{dio4}{50}{3}{258}{188}{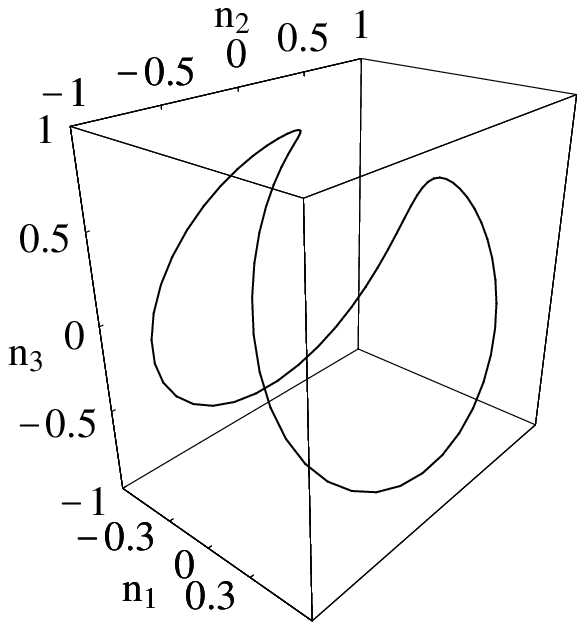}{5.2}{7.5}{
Spin components $n_{\alpha}$ of the orbit Q$_1$ in the spin-boson system 
\eq{sbclass} at $\kappa=0.30$. Other parameters as in \fig{dishap}.}
\end{minipage}

\subsection{Two-dimensional quantum dot with Rashba spin-orbit interaction}
\label{secrashba}

We finally consider a two-dimensional electron gas in a semiconductor 
heterostructure, laterally confined to a quantum dot by a harmonic potential. 
It is modelled by the quantum Hamiltonian
\be 
  \hat{H} = \frac12\,(\hat{p}_x^2 +\hat{p}_y^2) 
          + \frac12\, (\omega_x^2 \hat{x}^2 + \omega_y^2 \hat{y}^2) 
          + 2 \kappa  (\hat{s}_2 \hat{p}_x - \hat{s}_1 \hat{p}_y)\,;
\label{qham} 
\ee
here we put the effective mass of the electrons to be $m^*=1$. The 
semiclassical treatment of this system has been presented recently in 
\cite{plet}. The classical symbol of the quantum Hamiltonian (\ref{qham})
\be
H = \frac12\,(p_x^2+p_y^2)+\frac12\,(\omega_x^2x^2+\omega_y^2y^2)+
    \kappa\,(n_2p_x-n_1p_y)
\label{hamrash}
\ee
was considered and the corresponding (semi-)classical equations of motion 
were studied there. Two analytic periodic solutions $A_{x}^{\pm}$ and 
$A_{y}^{\pm}$, as well as four numerical solutions $D_{x1}^{\pm}$, 
$D_{x2}^{\pm}$, $D_{y1}^{\pm}$, and $D_{y2}^{\pm}$, were found and 
discussed in \cite{plet}. 

We present in \tab{tabrash} the Maslov indices of the twelve shortest 
periodic orbits of this system, calculated at the same parameter values as 
in \cite{plet}. 
Note that the Hamiltonian (\ref{hamrash}) describes a system which is 
effectively three-dimensional. Therefore, loxodromic blocks occur in the 
monodromy matrix, as well as transitions from a loxodromic block into two 
elliptic blocks without change in the Maslov index \cite{sugita}. We should 
also mention that for the calculation of the Maslov indices quoted in 
\tab{tabrash} we have used different Darboux representations to avoid the 
problem of crossing the pole of projection. Thus, for $D_{x1}^{\pm}$ and 
$D_{x2}^{\pm}$ we have chosen 
\be
n_1 = -q_z \sqrt{2 - p_z^2 - q_z^2}\,, \qquad   
n_2 = -(p_z^2 + q_z^2 -1)\,, \qquad
n_3 = p_z \sqrt{2 - p_z^2 - q_z^2}\,,
\ee
which corresponds to the pole of projection located at $(0,-1,0)$, while
for $D_{y1}^{\pm}$ and $D_{y2}^{\pm}$ we have projected from the point
$(1,0,0)$
\be
n_1 = p_z^2 + q_z^2 -1\,, \qquad   
n_2 = q_z \sqrt{2 - p_z^2 - q_z^2}\,, \qquad
n_3 = -p_z \sqrt{2 - p_z^2 - q_z^2}\,.
\ee
Hereby we have put $q_z \equiv Q_s$ and $p_z \equiv P_s$.

\Table{tabrash}{16.6}{
\begin{tabular}{|c|c|c|c|c|c|r|c|} \hline
orbit & blocks & ${\rm sign}( s_1 , s_2)$ & $m$ & $\sigma_{\mathrm{av}}$ & 
$\widetilde{m}$ & $\widetilde{\sigma}_{\mathrm{av}}$ & $\sigma$   
\\ \hline\hline
$A_x^{\pm}$    & ell, ell &  $-$ , $-$       & 1 & 2 & 3 & $-2$ & 4 \\ \hline
$D_{x1}^{\pm}$ & hyp, ell &  ~~ , $-$        & 2 & 1 & 3 & $-1$ & 5 \\ \hline
$D_{x2}^{\pm}$ & ell, ell &  $-$ , $-$       & 0 & 2 & 2 & $-2$ & 2 \\ \hline
$A_y^{\pm}$    & lox      &                  & 3 & 0 & 3 & 0    & 6 \\ \hline
$D_{y1}^{\pm}$ & lox      &                  & 2 & 0 & 2 & 0    & 4 \\ \hline
$D_{y2}^{\pm}$ & hyp, ell &  ~~ , +          & 2 & 1 & 2 & +1   & 5 \\ \hline
\end{tabular}
}{~Stabilities, Maslov indices and their ingredients of the shortest orbits
in the Rashba Hamiltonian \eq{hamrash}.
}

\section{Summary}
\label{secsum}

In this paper we have taken the point of view of practitioners of the
semiclassical periodic orbit theory. We have formulated a simple
calculational recipe for the calculation of Maslov indices for isolated
periodic orbits that is canonically invariant and does not require the 
use of the orbits' intrinsic coordinate systems. Our work was inspired by 
two recent formulations \cite{sugita,murat} which are theoretically very 
thorough but both have left some practical questions unanswered. We have 
given unique and practicable definitions of the stability angle $\chi$ and
the winding number $m$, which are the main ingredients of Sugita's formula 
\eq{main} for the Maslov index, and tested them for an integrable and
various non-integrable systems. We have found that this formula leads to 
identical results with the method of Wintgen et al \cite{wint} and the
method of Creagh et al \cite{crliro}. An alternative definition of 
stability angle and winding number, using a different phase convention, 
allowed for a direct relation to the decomposition \eq{munu} given in 
\cite{crliro} and lead us to formulate some empirical rules which are 
useful for the classification of periodic orbits in connection with 
complicated bifurcation scenarios. These rules could also be verified in a 
novel sequence of periodic orbits that we have found in the H\'enon-Heiles
system to generate from a tangent bifurcation occurring near the saddle
energy. Their shapes are so entangled that the use of their intrinsic
coordinate systems needed in the methods of \cite{wint,crliro} was 
numerically not stable enough to yield unique Maslov indices. The present
method gives unique results as long as these orbits are not too unstable
($|$tr$M_{\rm red}|\siml 40$), thus demonstrating the practical strength
of this method.

We do not claim to have established any fundamentally new insights here.
As a matter of fact, some of our steps and observations have been hinted
at before in the literature \cite{liro,crliro,sugita,wnrobb}. Our aim
was rather to clarify some practical aspects and to define an easy-to-use 
but canonically invariant method for the calculation of Maslov indices,
applicable to the most general type of Hamiltonian systems including 
spin degrees of freedom. We believe to have reached this goal and hope 
that our method turns out to be useful also for other practitioners.

\newpage

\section*{Acknowledgements}

\ms
\noindent
We are gratefully indebted to Stephen Creagh, Jonathan Robbins and Ayumu 
Sugita for helpful discussions, and to Maurice and Serge de Gosson, J\"org
Kaidel and Klaus Richter for their stimulating interest. We also appreciate 
a clarifying exchange with Paolo Muratore-Ginanneschi. We acknowledge the 
use of numerical codes written by Kaori Tanaka \cite{lamp} and Christian 
Amann \cite{aman} for searching periodic orbits and calculating their 
stabilities. This work was supported by the Deutsche Forschungsgemeinschaft.

\end{document}